\documentclass[12pt]{iopart}
\usepackage{graphicx}
\usepackage[T1]{fontenc}
\usepackage{lmodern}
\def \gmt {Ge$_{1-x}$Mn$_x$Te}

\usepackage{iopams}  
\begin{document}

\title[Calculations of magnetic anisotropy energy of \gmt]{DFT calculations of
magnetic anisotropy energy of \gmt\ ferromagnetic semiconductor}

\author{A {\L}usakowski$^1$, P Bogus{\l}awski$^{1,2}$ and T Story$^1$}

\address{$^1$Institute of Physics, Polish Academy of Sciences, Al. Lotnik\'{o}w
32/46, 02-668~Warsaw, Poland\\
$^2$ Institute of Physics, University of Bydgoszcz, ul. Chodkiewicza 30,
85-072~Bydgoszcz, Poland}
\ead{lusak@ifpan.edu.pl}
\vspace{10pt}
\begin{indented}
\item[]February 2015
\end{indented}

\begin{abstract}
Density functional theory (DFT) calculations of the
energy of magnetic anisotropy  for  diluted ferromagnetic semiconductor \gmt\ 
were performed using OpenMX package with
fully relativistic pseudopotentials. The influence of hole concentration
and magnetic ion neighbourhood on magnetic anisotropy energy is presented.
Analysis of microscopic mechanism of magnetic anisotropy is provided, in
particular the role of spin-orbit coupling, spin polarization and spatial
changes of electron density are discussed.
The calculations are in accordance with the experimental observation of 
perpendicular magnetic anisotropy  in rhombohedral \gmt\ (111) thin
layers.
\end{abstract}

%
%
%
%
%

\section{Introduction}
Microscopic understanding of anisotropic properties of any ferromagnet
constitutes a very challenging theoretical task and is very important 
from the point of view of possible applications. Particularly, in the case of
ferromagnetic semiconductors, a deep understanding of the mechanisms
responsible for the behavior of magnetization becomes crucial because 
a number of semiconductor properties may be changed by external factors. It was
established that direction of the easy axis of magnetization can be changed by
changing the density of free carriers or by applying a strain.
The carrier concentration can be varied by applying gate voltage to the layer,
and strain in a film can be imposed by cementing it to a piezoelectric
actuator.
These and other related phenomena are described in a recent review
\cite{dietl_ohno}.

Previous theoretical descriptions of magnetic anisotropy 
in semiconductors 
were based either on the mean-field Zener model, 
or on the concept of single ion anisotropy. 
The Zener model was applied to III-V and II-VI diluted ferromagnetic
semiconductors \cite{dietl_ohno1,dietl_ohno2}.
In this approach, it is assumed that 
the nonzero macroscopic magnetization of magnetic ions leads, via $sp-d$
exchange interaction, to the spin splitting of band states. 
Because of the presence of spin-orbit (S-O) interaction, 
the band structure, and thus the energy of carriers populating band
states, depend on the direction of magnetization. 
In this model the magnetic anisotropy is due to
the simultaneous presence of two factors, i.e. the finite 
concentration of free carriers, and the non-vanishing S-O coupling 
in the host crystal. 
For insulating crystals, or for the vanishing band S-O 
interaction, a vanishing magnetic anisotropy is predicted. 
(Note that the limit of isolated impurities is beyond the scope 
of the model.)
The Zener model explained, at least semiquantitavely, a number of
experiments \cite{dietl_ohno1, stefanowicz, sawicki1,sawicki2}.\\
The second approach is based on the concept of single ion 
anisotropy. 
The anisotropic properties of a macroscopic crystal are 
brought about by anisotropic magnetic properties of magnetic 
$3d$ or $4f$ ions subject to ligand electric fields of 
the crystal environment.
In II-VI and IV-VI semiconductors, the Mn$^{2+}$ ions 
in the crystal retain to some extend their atomic-like character. 
In the absence of the S-O coupling, 
according to the Hund's rule five electrons on $3d$
shell form an orbital singlet with the total angular momentum $L=0$. 
This system does not interact with the crystal field, 
and the ground state of the ion is sixfold degenerate since the total spin
$S=5/2$.  
Actually, however, the Hund's rule is an approximation only, 
because due to relativistic S-O effects there is a coupling between 
$^6S$ and the excited $^4P$ states of $3d^5$ configuration. 
Consequently, the ground state is no longer a pure $S$ state, and the coupling
of Mn with its crystal surrounding becomes possible. 
This leads to the ground state splitting, and Mn can be described by the
$6\times 6$ matrix of the effective spin hamiltonian \cite{abragam}.
For example, the spin hamiltonian of Mn$^{2+}$ in a cubic environment 
is proportional to $S_x^4+S_y^4 +S_z^4$. 
This effect is experimentally observed as a fine structure splitting of the
electron paramagnetic resonance (EPR) spectra \cite{abragam}. 
During the last five decades various calculations of this effect were presented,
see for example Refs.~\cite{sharma1,sharma2,sharma3,vanluan}. 
However, in these works the virtual, excited states taken into account in the
perturbation treatment had the same number of electrons on $3d$ orbitals as the
ground state, i.e., five.

This constraint was lifted in 
another approach to the single ion anisotropy 
proposed in Ref.~\cite{lusakowski2}. In this work,  
the system consisted of the Mn$^{2+}$ ion and band carriers, 
and in the perturbation calculations of the effective spin hamiltonian 
the number of electrons on the 
excited $3d$ states could be different than five. 
In other words, the virtual transfer of electrons between the $3d$ shell and
band states was allowed. 
This idea was previously widely used in the calculations 
of $sp-d$ kinetic exchange integral in semiconductors \cite{larson, dietl_sliwa},
or in the calculations of exchange coupling between magnetic ions \cite{kacman}.
 
Within this approach it was possible to study the influence of 
both carrier concentration and alloy disorder on the ground state splitting. 
The method succesfully explained the EPR measurements of Mn ion 
in strained PbTe layers \cite{lusakowski2},
the single ion magnetic anisotropy in Pb$_{1-x-y}$Sn$_{x}$Mn$_{y}$Te mixed
crystals \cite{lusak_ssc},
the magnetic specific heat in Pb$_{1-x}$Mn$_{x}$Te \cite{lusakowski3}, as well
as that of 
PbTe doped with Eu$^{2+}$ and Gd$^{3+}$ \cite{lusakowski4}. 

In the above theories of single ion magnetic anisotropy the manganese spin is
treated as a quantum mechanical object. In particular, the nonvanishing S-O 
coupling for $3d$ electrons is necessary to produce a nonzero 
ground state splitting, or single ion anisotropy of Mn. 
The band S-O coupling is of minor importance in the sense 
that even for the zero band S-O interaction we
get the splitting, although much smaller than in the case of 
the actual S-O coupling. 
On the other hand, it is also possible to propose a model 
of single ion anisotropy \cite{lusak_app} in which the manganese spin, like in
the Zener
model, is treated as a classical Heisenberg vector interacting with band
carriers via $sp-d$ interaction,  but of course, in this case the nonzero band
S-O coupling is necessary. In this model, a strong dependence of magnetic
anisotropy constants on the concentration of free holes was predicted and
qualitatively confirmed experimentally \cite{eggenkamp} in
Sn$_{1-x}$Mn$_{x}$Te. 

In the present paper we study magnetic anisotropy in \gmt\  ferromagnetic
semiconductor by
first principle calculations. A natural question arises what, concerning the
magnetic anisotropy, significantly new information may be obtained applying
this method? Certainly, using DFT calculations it is not possible to obtain
directly the ground state splitting or the effective spin hamiltonian for the
Mn
ion. This is not related to the accuracy problem or the limitations in
computational resources but to the fact that DFT calculations determine 
the charge density in the ground state of the system. On the other
hand, as it is shown below, this approach allows to
investigate microscopic mechanisms responsible for the anisotropy,
in particular the role of S-O coupling, of the range
of spin polarization of atoms surrounding the magnetic ion,
and of the influence of crystal disorder.
Although we concentrate exclusively on \gmt\,
we think that the general conclusions may be useful for other
ferromagnetic semiconductors.

GeTe is a nonmagnetic narrow gap IV-VI semiconductor, which
has rhombohedral crystal symmetry with the angle characterizing
the unit cell $\alpha \approx 88.35^o$ at room temperature.
In addition, the Ge and Te sublattices
are displaced relative to each other along the [111] direction
by a vector $\tau a_0(111)$, where the lattice constant
$a_0 =5.98$~\AA\ and $\tau \approx
0.03$ \cite{fukuma0,przyby}; this displacement causes GeTe to be
a unique ferroelectric.
These features also characterize magnetic \gmt\ mixed crystals
for moderate manganese contents $x < 0.2$.
\\
In IV-VI semiconductors, Mn ions are incorporated as Mn$^{2+}$
with the $3d^5$ electron configuration. They introduce
local magnetic moments \cite{story_swuste} and remain
electrically neutral.
\gmt\  diluted magnetic semiconductor
exhibits ferromagnetic transition with the Curie
temperature $T_C(x)$ up to 190~K \cite{lechner, fukuma}. 
In these mixed crystals ferromagnetism is induced by quasi-free carriers.
The simultaneous presence of ferromagnetic and
ferroelectric order (multiferroicity) makes this compound very interesting for
possible applications,
in particular as a magnetic phase-change material, or a
spintronic system with electrical polarization controlled spin
splittings of band states \cite{picozzi}.

Recently, magnetization and ferromagnetic resonance (FMR) measurements were
performed on both monocrystalline layers of \gmt\  grown on the BaF$_2$ (111)
substrate and on polycrystalline, layered Ge$_{0.9}$Mn$_{0.1}$Te
microstructures \cite{przyby,fukuma,knoff1,knoff2,fukuma1}.
The experiments probed magnetic anisotropy energy (MAE) in
the ferromagnetic state.
One of the striking results is that in monocrystalline layers with the manganese
content of the order of 10~at.~\%, the easy axis of magnetization is oriented
along the [111] crystallographic direction perpendicular to the layer,
while the usual in-plane easy axis due to dipolar interactions
(shape anisotropy) is observed in polycrystalline microstructures and is
partially recovered in
strained monocrystalline \gmt\  layers upon annealing.
In the layers with the Mn content higher than 20~at.~\%, for which the X-ray
diffraction
studies reveal the cubic (rock-salt) structure, the usual in-plane easy axis is
observed.

Motivated by the experimental results we study microscopic mechanisms of
magnetic anisotropy and  the decisive factors that determine MAE.
We mainly report the results obtained for rhombohedral crystals
characterized by $a_0 =5.98$~\AA, $\alpha = 88^{o}$  and $\tau \approx 0.03$,
and low Mn concentrations.
Most of the
calculations were performed for $2\times 2\times 2$ supercells containing 64
atoms
with one Ge replaced by Mn atom, which corresponds to
approximately $x=0.03$ in \gmt\ mixed crystal.
For comparison, we present also some results for the 216 atoms
$3\times 3\times 3$ supercell containing one Mn atom, which corresponds to
$x\approx 0.01$.
These Mn concentrations are much smaller than those in the
samples studied experimentally, and this is the first reason why a quantitative
comparison
with experiment is not possible. The second  reason is also important:
preliminary calculations clearly indicated that MAE
strongly depends on the microscopic disorder present in an alloy.
This disorder is inevitable in mixed crystals due to both the different ionic
radii of ions
constituting the crystal, and to the random occupation of lattice sites.
Consequently, in an actual crystal each Mn ion is in a different surrounding.
In general, these surroundings have no symmetry, what leads
to different directions of the easy axis of magnetization for different Mn ions.
Below, an example of the dependence of MAE on the nearest
neighbourhood of Mn ion is given.
However, a detailed analysis of influence of the microscopic disorder
is beyond the scope of the paper.
That is why geometry optimization of the supercells typically was neglected.  

The paper is organized as follows.
After presenting the  technical details of calculations in Section II,
in Section III we analyze the dependence of MAE on the concentration
of free holes and on the manganese ion's neighbourhood.
Section IV contains analysis of various factors that determine MAE.
We discuss the role of spin-orbit coupling, of spin polarization of free
carriers in the supercell, and of the changes
of the electron density caused by the changes of manganese spin's direction.
Section V summarizes the paper.
A few preliminary results were presented
in a short conference communication \cite{lusak_jasz}.

\section{Technical details of calculations}

The calculations were performed with the open-source OpenMX
package \cite{openmx} with fully relativistic pseudopotentials.
The Local Density Approximation together with the Ceperly-Alder
\cite{CA}  exchange-correlation functional was used.
The inclusion of spin-orbit  interaction is crucial for calculation
of energy of magnetic anisotropy, which disappears when the spin-orbit
interaction is omitted.

\subsection{Pseudopotentials}
The pseudopotentials distributed with OpenMX are generated assuming
16 valence electrons for Te.
Because of the very high accuracy demanded by the studied problem,
large sizes of supercells and very high density of grids in the
k-space,
it was necessary to use pseudopotentials with lower numbers of valence
electrons.
These were  generated with the program ADPACK, contained in the OpenMX package.
We generated fully relativistic pseudopotential with the $5s^25p^4$
configuration, $i.e.$, with 6 valence electrons, for Te,
and with the $4s^24p^2$ configuration for Ge.
The pseudopotentials were calculated using Troullier-Martins
\cite{TM} algorithm with Kleinmann-Bylander factorization \cite{KB}.
For Ge, the cutoff radii for $4s$ ($4p$) pseudopotentials were 1.80 (1.70)
atomic units (a. u.). For Te, the cutoff radii were 2.40 (2.20) a.~u. for $5s$
($5p$) pseudopotentials.
The radial cutoffs for pseudoatomic orbitals were 7.5 and 8.0 a.~u.
for Ge and Te, respectively.
Furthermore, for Ge and Te we used the minimal wave functions basis,
$i. e.$, $s1p1$ in the notation explained in Refs. \cite{ozaki1,ozaki2},
which means that in the wave function basis for Te and Ge atoms one $s$ and
three
$p$ orbitals
($p_x$, $p_y$ and $p_z$) were included. The number of orbitals was doubled
due to the inclusion of electron's spin.
These pseudopotentials and the basis set lead to equilibrium lattice parameters
$a_{th}$=6.0 \AA, $\alpha_{th}=89.3^o$ and $\tau_{th}=0.019$,
which are close to the experimental values.
For comparison, the original pseudopotentials distributed with
OpenMX package give $a_{th}$=6.09 \AA, $\alpha_{th}=89.5^o$ and
$\tau_{th}=0.014$.

For Mn, we used the pseudopotential distributed with OpenMX
with 15 valence electrons ($3s^23p^63d^54s^2$ configuration),
together with pseudoatomic basis $s3p2d1$.
This choice increases the number of valence electrons,
but since there is only one Mn atom in the supercell,
the computational time is practically not affected.

Energy levels of atomic shells with $l\ge 1$ are always split by the S-O
interaction. The energy difference between split levels, $\Delta E = \Delta^0E$,
where $\Delta^0E =
E(j=l+1/2)-E(j=l-1/2)$, is directly related to the S-O strength.
Usually, atomic pseudopotentials for $j=l\pm 1/2$ are calculated
assuming that the values of these splittings are those for a real atom,
and such $j$-dependent pseudopotentials are called fully relativistic.
On the other hand, by performing appropriate average \cite{bachelet}
over $j$-dependent pseudopotentials, it is possible to obtain the so-called
scalar relativistic pseudopotentials for which $\Delta E = 0$.
For these two cases, $\Delta E=\Delta^0E$ and $\Delta E= 0$, the factor
$f_{so}$ describing in the following the S-O strength is $f_{so}=1$ and
$f_{so}=0$, respectively.
OpenMX package offers the possibility to generate $j$-dependent
pseudopotentials for arbitrary $f_{so}$ factors,
for which $\Delta E = f_{so}\Delta^0E$. This mimics
the variation of the S-O interaction strength for a given shell of an atom.
To study the influence of the S-O interaction on MAE, we generated
pseudopotentials for Ge and Te for a few values of $f_{so}$, while
for Mn we always used the pseudopotential with $f_{so}=1$,
corresponding to the actual atomic S-O strength.\\

\subsection{Constraints of the Mn spin direction}
An important feature of OpenMX is the possibility to
constrain directions of spin polarization for atoms.
This is done by adding a properly constructed harmonic
potential to the total energy functional. To introduce  notation
which is used below, we briefly sketch
this construction \cite{openmxtechnot}.
In relativistic calculations the wave function is a two component spinor
\begin{equation}
 |\psi_{\nu}>=\left(\begin{array}{c}\phi_{\nu}^{\alpha}\\
                     \phi_{\nu}^{\beta}
                    \end{array}
\right),
\label{eq1}
\end{equation}
and the electron density operator $\widehat{n}$
is a $2\times 2$ matrix

\begin{equation}
\widehat{n}=\sum f_{\nu} |\psi_{\nu}> <\psi_{\nu}|,
\label{eq2}
\end{equation}
where $f_{\nu}$ is the occupation number.
By projecting this matrix on
orbitals of a given atom one obtains a matrix
$\widehat{N}$, which can be diagonalized by an unitary matrix $U(\theta,\phi)$
dependent on two angles, $\theta$ and $\varphi$:
\begin{equation}
U(\theta,\varphi)\widehat{N}U^{\dagger}(\theta,\varphi) =
\left(\begin{array}{cc}N_{\uparrow}&0\\0&N_{\downarrow}\end{array}\right).
\label{eq3}
\end{equation}
The angles $\theta$ and $\varphi$ define both the direction of total spin
and the difference $\pi=N_{\uparrow}-N_{\downarrow}$, which is the spin
polarization of a given atom.\\
To constrain spin of the atom to the direction
defined by angles $\theta_0$ and $\phi_0$, the following term
is added to the total energy functional:

\begin{equation}
 E_{cs}=v {\rm Tr}\left(\left(\widehat{N}-\widehat{N}_0\right)^2\right),
\end{equation}

where
\begin{equation}
 \widehat{N}_0=U^{\dagger}(\theta_0,\varphi_0)\left(
\begin{array}{cc}N_{\uparrow}&0\\0&N_{\downarrow}\end{array}
\right) U(\theta_0,\varphi_0).
\end{equation}
If after the selfconsistent procedure the spin direction is
the same as the demanded one, the energy connected with this term
is zero. Indeed, after calculations this energy was always
negligible compared to other energies of the system,
and the final angles $\theta$ and
$\varphi$ describing the spin directions were practically the same
as those demanded provided the spin polarization of atoms was not negligible.

In most calculations, directional constraints were applied only to Mn.
Analysis of the results shows that the spins of Ge atoms are
parallel to the spin of Mn, and the directions of Te spins are opposite.
Moreover, the calculated MAE does not depend significantly on whether
the constraint was applied to the Mn spin only,
or to all atoms in the supercell.
We conclude therefore that the
direction of Mn spin governs the directions of Ge and Te spins.

In all calculations, the Mn spin is rotated in (1$\bar{1}$0) plane and the
angle $\theta$ is measured in this plane from the [111] direction.
Such a rotation corresponds to the situation frequently
studied in electron ferromagnetic resonance experiments,
when the external magnetic field rotates in this plane.
\gmt\  on BaF$_2$ substrates grows along the [111] direction,
thus $\theta=0$ and $\theta = 90^{\circ}$ correspond to the Mn spin
perpendicular and parallel to the plane of growth, i.e., oriented
along the [111] and[11$\bar{2}$] crystallographic directions, respectively.

\subsection{Other input parameters}

The calculations were performed in part for nonzero densities of free
holes $p$, because actual \gmt\  samples are $p$-type due to very high
concentrations of native defects (Ge vacancies).
This additional positive charge was compensated by a uniform negative
background. The hole concentrations assumed in calculations were in the
interval $0\le p \le 5\times 10^{21}$cm$^{-3}$.
To speed up the convergence, particularly in the case
of nonzero hole concentrations, the calculations were performed assuming
the electron temperature 300~K.

The integrals in numerical calculations are approximated by sums over
discrete sets of points. Thus, it is important to properly choose the grid
for the system in the real and the reciprocal space to secure an optimal
balance between the accuracy and the computational time.
Convergent results are obtained
for the length of the grid in real space approximately equal to 0.15~\AA.
Much more important is the number of integration
points in the Brillouin zone giving convergent results,
which, as it is discussed below, strongly depends on the concentration of holes.

\subsection{Angular dependence of MAE}
Group theory predicts a general dependence of MAE
on the direction of magnetization.
The knowledge of this dependence enables significant
saving of computer time.
For the $O_h$ crystal symmetry, the first nonvanishing terms in MAE
are of the fourth order in
magnetization, or in the total spin components $S_i$, $i=x,y,z$.
When the symmetry is lowered to $C_{3v}$ for rhombohedral \gmt,
the lowest order terms are proportional to $S_i^2$ and $S_i^4$, and MAE can be
written as
\begin{equation}
\label{eq6}
 E_A(\theta) =  a_2\cos 2\theta + b_2\sin 2\theta +
a_4\cos 4\theta + b_4\sin 4\theta  +c,
\end{equation}
where $\theta$ is the angle between the direction of the Mn spin placed in
(1-10) plane and the [111] crystallographic direction,
and $c$ is the isotropic part chosen in such a way that $E_A(\theta=0)=0$.

In most cases, the rhombohedral angle
$\alpha= 88^o$ and the displacement $\tau = 0.03$ were assumed.
For these parameters the first term in Eq.~(\ref{eq6}) dominates,
and MAE may be written in the form
\begin{equation}
\label{eq7}
 E_A(\theta) =  a_2(\cos 2\theta -1).
\end{equation}
The remaining terms in Eq.~(\ref{eq6})
are important when the crystal structure is close to the
perfect cubic, or when two or more  Mn ions are placed in the supercell.
The validity of the fitting formula, Eq.~(\ref{eq7}), is
fully confirmed by the obtained numerical results.
Moreover, the correctness of Eq. (\ref{eq7}) for rhombohedral
\gmt\ is confirmed by ferromagnetic
resonance measurements \cite{przyby,knoff1a}, which also lead to
the
conclusion that the terms proportional to the fourth order
in magnetization are very small.

\section{Factors that determine MAE}
\subsection{Band structure of G\lowercase{e}T\lowercase{e}}

\begin{figure}[t]
\includegraphics*[width=\textwidth]{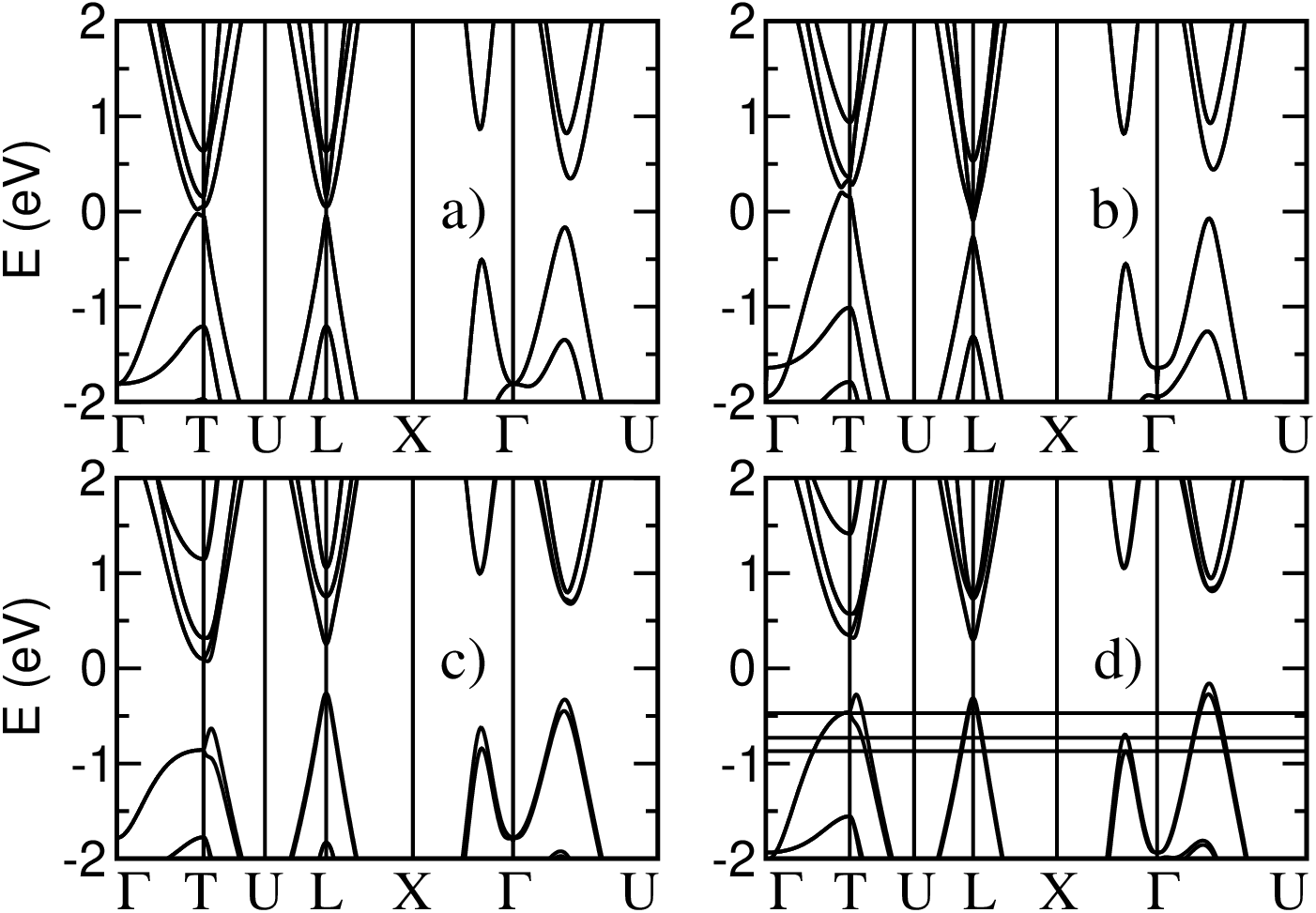}
\caption{%
GeTe band structure for different crystallographic structures:
(a) $\alpha=90^o$, $\tau = 0$, (b) $\alpha=88^o$, $\tau = 0$, (c) $\alpha=90^o$,
$\tau = 0.03$, (d) $\alpha=88^{o}$, $\tau = 0.03$. The horizontal lines at (d)
represent positions of the Fermi level for hole concentrations
$10^{21}$cm$^{-3}$, $3\times 10^{21}$cm$^{-3}$, and $5\times 10^{21}$cm$^{-3}$,
respectively.}

\label{fig1}
\end{figure}

We first discuss the impact of the crystal distortions on
the band structure of pure GeTe.
Figure \ref{fig1} shows the band structure for four
sets of structural parameters, namely
(a) $\alpha=90^o$, $\tau = 0$, (b) $\alpha=88^o$, $\tau = 0$,
(c) $\alpha=90^o$, $\tau = 0.03$, and (d) $\alpha=88^{o}$, $\tau = 0.03$.
In the first case the crystal symmetry is $O_h$, and it is lowered
to $D_{3d}$ for (b), and to $C_{3v}$ for both (c) and (d).

When the symmetry is lower than $O_h$, the $T$ and $L$ points in
the Brillouin zone are nonequivalent. They are defined by
$T=\frac{1}{2}({\bf b_1} + {\bf b_2} + {\bf b_3})$  and $L=\frac{1}{2}{\bf
b_1}$,
respectively, where ${\bf b_1}$, ${\bf b_2}$, ${\bf b_3}$ are the primitive
vectors of the reciprocal lattice. Other points in the Brillouin zone of a
rhombohedral lattice are defined in Ref.~\cite{falicov}.

As it follows from Fig.~\ref{fig1}, the impact of the rhombohedral
distortion on the band structure is much weaker than that of the
internal displacement of the Ge and Te sublattices, since
the latter opens the band gap of about 0.5 eV at both $L$ and $T$ points.
One can also observe that the decrease of the angle $\alpha$
leads to the decrease of band
energies in the vicinity of the $L$ point compared to those in
the vicinity of $T$, while the increase of parameter $\tau$ has the opposite
effect.

The band structures shown in Fig. 1 are calculated for wavefunction
basis s1p1. Qualitatively, the same results are obtained for enlarged
bases s2p2 and s3p3.

It is known that the LDA calculations underestimate the energy gap.
Due to this effect, GeTe in the $O_h$ structure is found to be
semimetallic with zero energy gap.
This is also the case when the rhombohedral distortion is accounted for
but the internal displacement $\tau$ is neglected, Fig.~\ref{fig1}.
For the present work it is important that the system
has a non-vanishing energy gap for the used crystal parameters,
close to the experimental ones,
$\alpha=88^{{\rm o}}$ and $\tau=0.03$.
This enables a comparison of magnetic anisotropy in
insulating and conducting crystals, and an analysis of the role
of free carriers.
For hole concentrations of the order of 10$^{21}$~cm$^{-3}$,
typical for \gmt\  \cite{przyby},
the Fermi level lies deep in the valence band and the precise
band structure in the vicinity of the band maximum is not important.

\subsection{Dependence of MAE on the hole concentration}

\begin{figure}
\includegraphics*[width=\linewidth,height=0.66\linewidth]{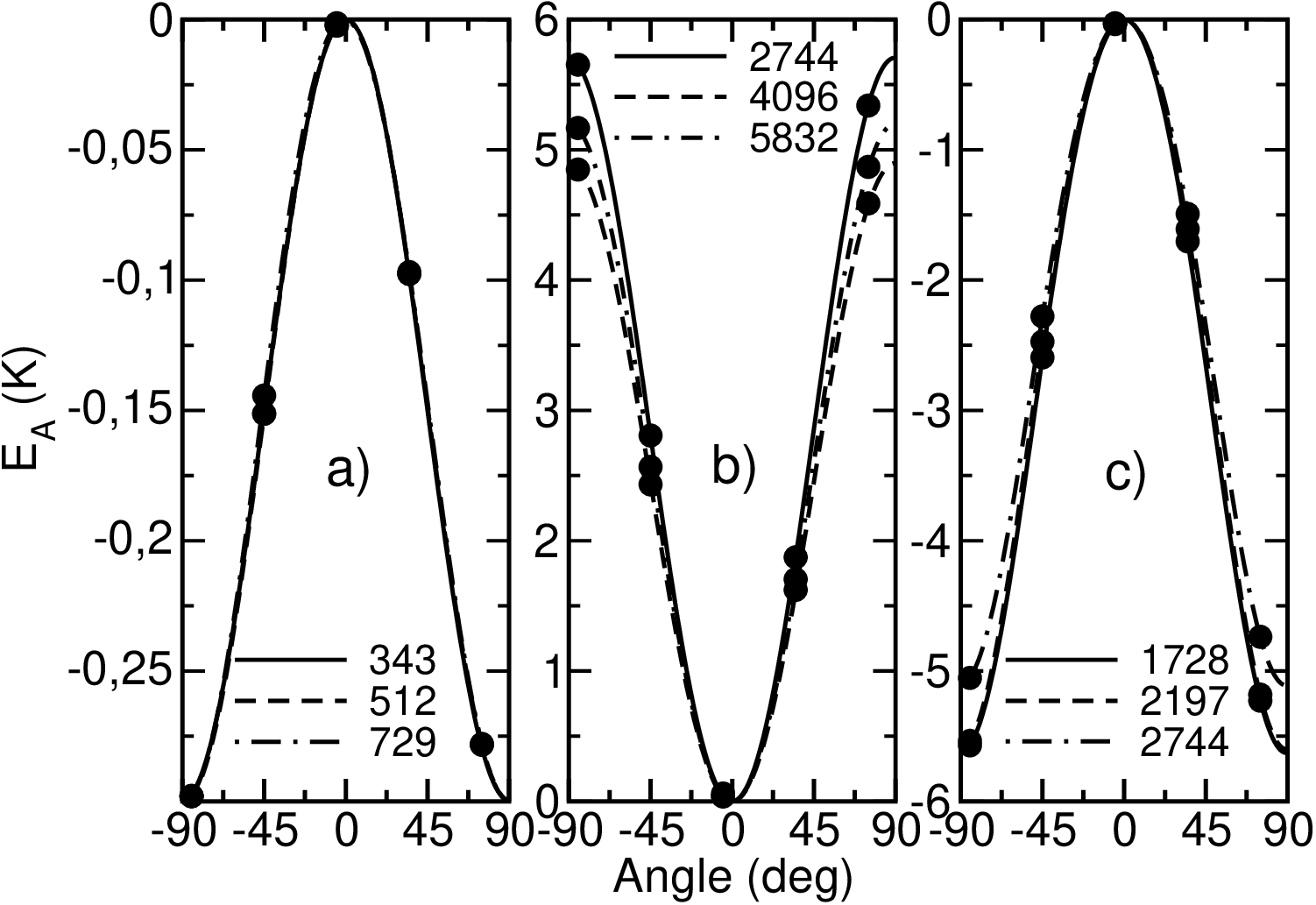}
\caption{%
Angular dependence of energy of magnetic anisotropy for the 
$2\times 2\times 2$
supercell and three different hole concentrations: (a) $p$=0,
(b) $p=2.5\times 10^{20}$ cm$^{-3}$,
and (c) $p=5\times 10^{21}$ cm$^{-3}$. The
numbers in legends are the numbers of grid points in the Brillouin zone.
}
\label{fig2}
\end{figure}

Figure \ref{fig2} presents the angular dependence
of MAE for the $2\times 2\times 2$ supercell containing
one Mn ion for 3 hole concentrations.
The calculated values are shown as points, and the curves are obtained
by fitting the results according to Eq. (\ref{eq7}), i.e.,
using one fitting parameter $a_2$.
In all cases, the angular dependencies of $E_A$ are symmetric
relative to $\theta=0$.

\begin{figure}
\includegraphics*[width=\linewidth,height=0.66\linewidth]{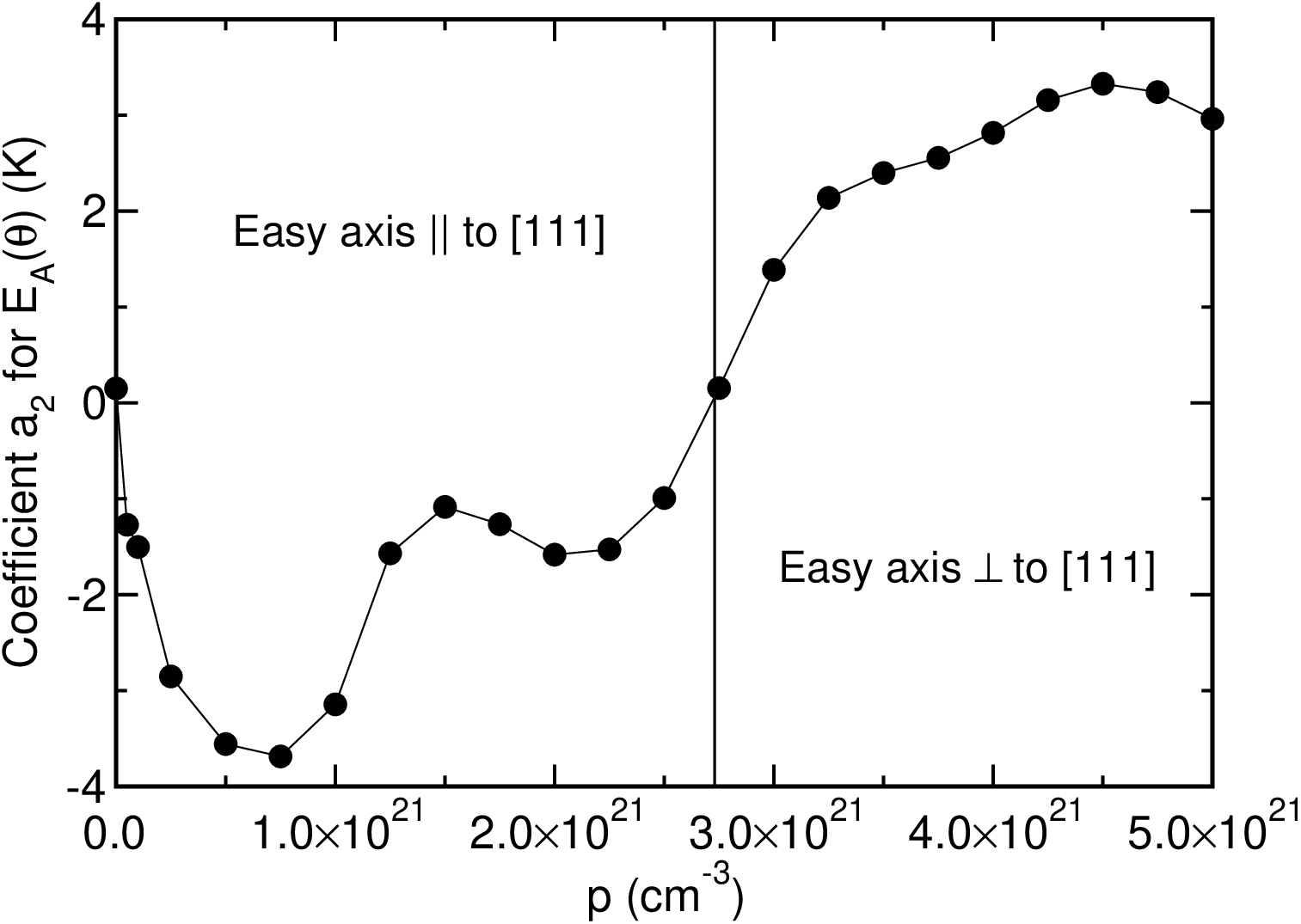}
\caption{%
The dependence of the coefficient $a_2$ of MAE (Eq. 2) on
the hole concentration.
The solid line is to guide the eye only. For $p>0$ all the calculations were
performed for 2744 integration points in Brillouin zone.
}
\label{fig3}
\end{figure}

The values of the coefficient $a_2$, calculated for several hole
concentrations $p$ ranging from zero to $5\times 10^{21}$~cm$^{-3}$,
are summarized in Fig. \ref{fig3}.
The obtained results show that both the absolute value of MAE
and its sign depend on the concentration of free carriers.
Indeed, the presence of free holes can enhance the magnitude of MAE
by more than one order of magnitude compared to the insulating case.
As discussed below, this is caused by a larger and more spatially
extended spin polarization around Mn in metallic samples.

Regarding the sign of MAE, three ranges of hole concentrations
can be distinguished. First, for vanishing or very small $p$,
the coefficient $a_2$ is positive, i.e., the equilibrium 
orientation of Mn spin is perpendicular to the [111] direction.
Second, in the window $1\times 10^{19}$ cm$^{-3}$ - $2.5\times 10^{21}$
cm$^{-3}$,
$a_2$ is negative. Finally, for higher $p$, $a_2$ is positive again.
Moreover, with the increasing hole concentration
a non-monotonic behavior is observed.

The impact of free carriers on magnetic properties 
reflects pecularities of the band structure. 
Changes of the direction of easy axis caused by changes in carrier
concentration were observed and analyzed 
for Ga$_{1-x}$Mn$_x$As \cite{chiba,sawicki3}.
Similarly, in Sn$_{1-x}$Mn$_{x}$Te,  the cubic anisotropy constant depends on 
the hole
concentration \cite{eggenkamp}.
Another effect related to the changes in carrier concentration
was observed in Pb$_{1-x-y}$Sn$_{x}$Mn$_{y}$Te diluted magnetic semiconductor, 
in which the Curie temperature depends non-monotonically 
on the hole concentration \cite{story1}.
The effect was explained 
by significant modifications of the RKKY interaction
caused by the two-band structure ($L$ and $\Sigma$ bands)
of the valence band \cite{story2}. 
The multiband structure of the valence band is characteristic 
for IV-VI compounds, and it is absent in $e.g.$ Ga$_{1-x}$Mn$_x$As. 
In the case under consideration, the number of valence bands 
occupied with holes depends on the Fermi energy, $E_F$. 
In Fig.~\ref{fig1}d we show the Fermi level position 
for three hole concentrations. 
For the highest concentrations, as much as four bands are 
occupied (at $T$, $L$, and those on both $X$-$\Gamma$ 
and $\Gamma$-$U$ directions). 
Each of those bands has an its own characteristic orbital composition, 
strength of the S-O coupling, $g$-factor, and coupling with Mn. 
For this reason, the non-monotonic dependence of MAE on 
the hole concentrations, shown in Fig. \ref{fig3}, 
can be related to progressive occupation of consecutive 
valence bands. 

\begin{figure}
\includegraphics*[width=\linewidth,height=0.66\linewidth]{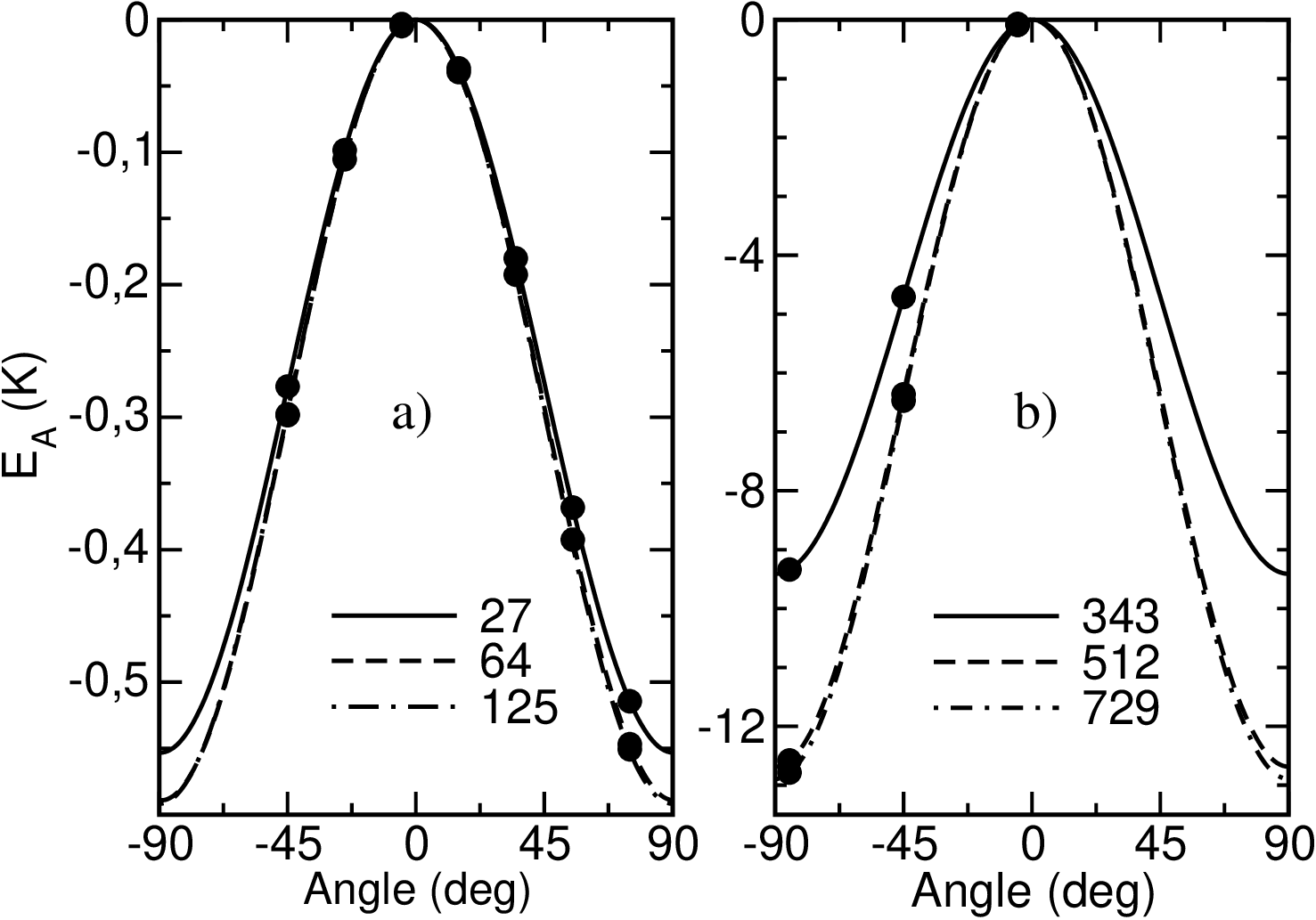}
\caption{%
Angular dependence of energy of magnetic anisotropy for the
$3\times 3\times 3$ supercell for two different hole concentrations:
(a) $p$=0  and (b) $p=5\times 10^{21}$ cm$^{-3}$. The
numbers in legends are the numbers of grid points in the Brillouin zone.
}
\label{fig4}
\end{figure}

Figure \ref{fig4} presents the results of calculations for
the $3\times 3\times 3$ supercell containing one Mn ion
for both the insulating and the metallic crystal
with $p=5\times 10^{21}$ cm$^{-3}$.
As it follows from the figure, the directions of easy axes
are the same as for the $2\times 2\times 2$ supercell.
However, the magnitude of $a_2$ is about twice higher for
the $3\times 3\times 3$ supercell in both insulating and metallic cases.
These results show that the Mn ions in the $2\times 2\times 2$
supercell cannot be treated as isolated, even in the insulating crystal.

Finally, also shown in Figs. \ref{fig2} and \ref{fig4} is the
convergence with respect
to the number of integration points (NIPs)
in the Brillouin zone. The convergence is easier to achieve for the
insulating system than for the metallic one.
In the former case, the obtained dependencies are practically
the same for three consecutive NIPs, $7^3$, $8^3$ and  $9^3$.
When the hole concentration is finite, the convergence is much slower.
Unfortunately, due to limitations in computational resources
we could not perform calculations for NIP greater than $14^3$
for the $2\times 2\times 2$ supercell  for all the considered 
cases. However, for
one selected concentration  $p=2.5 \times 10^{20}$~cm$^{-3}$, NIP was increased 
to
16$^3$=4096 and 18$^3$=5832, and the results shown in Fig. 2b confirm that a
satisfactory accuracy of about 5 per cent was achieved.

The increase of the supercell size leads to the so-called folding
of the Brillouin zone. Due to the folding,
the grid spacing obtained with NIP=$12^3$ used for
the $2\times 2\times 2$ supercell is the same as
NIP=$8^3$ used for the $3\times 3\times 3$ supercell, and both
correspond to the grid generated with NIP=$24^3$ for
the simple cubic unit cell.
From Figs \ref{fig2} and \ref{fig4} it follows that
the accuracy of the coefficient $a_2$ calculated for metallic
systems is about ten per cent, and less than one per cent
for insulating systems.
\\

\subsection{Dependence of MAE on the macroscopic 
 structural parameters $\alpha$ and $\tau$}

\begin{figure}
\includegraphics*[width=\linewidth,height=0.66\linewidth]{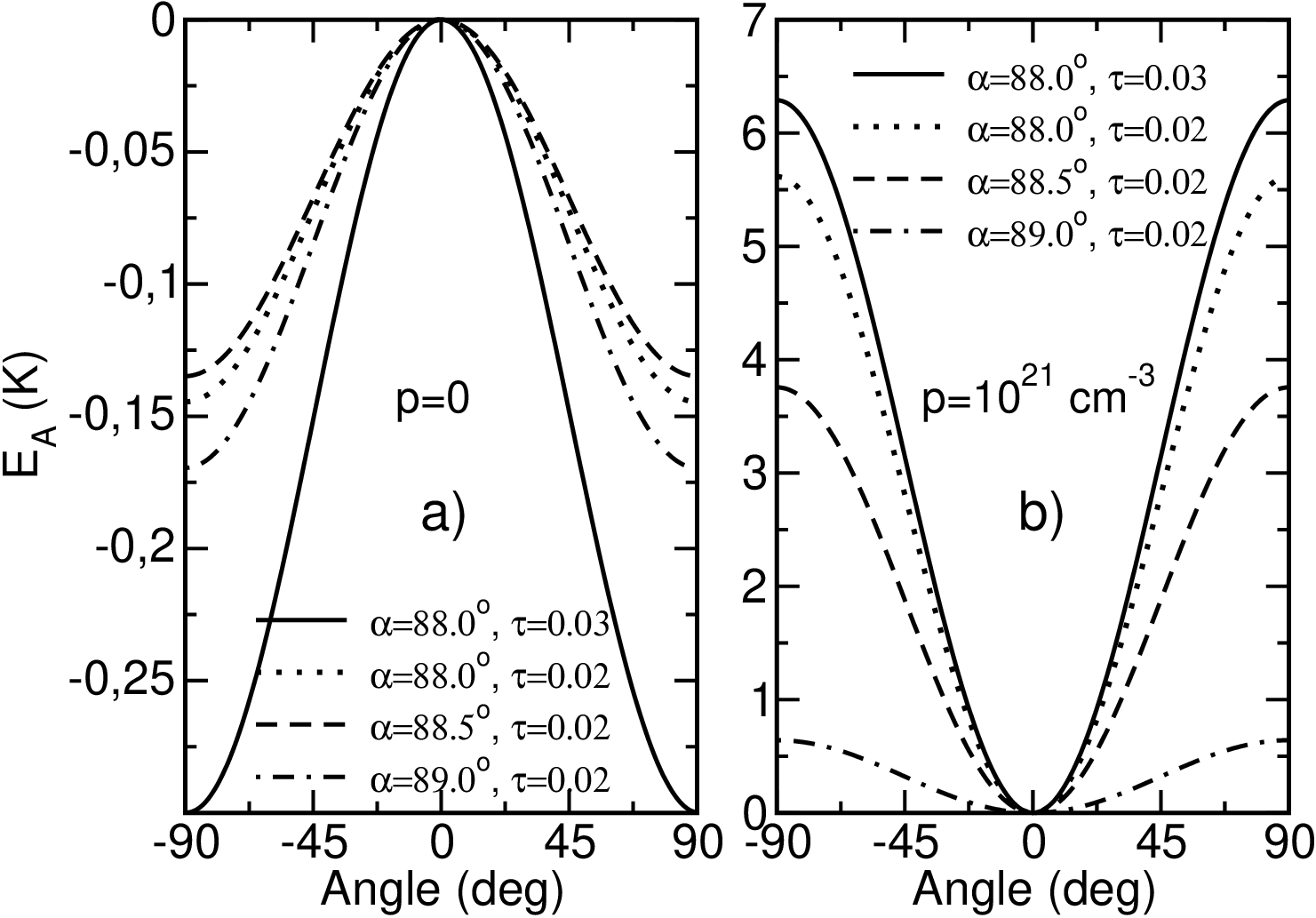}
\caption{%
Angular dependence of MAE for
the $2\times 2\times 2$ cell for (a) $p=0$, NIP=512 and (b) for  
$p=10^{21}$cm$^{-3}$, NIP=2744 for different crystal structure
parameters $\alpha$ and $\tau$.}
\label{fig5}
\end{figure}

Figure ~\ref{fig5} shows the angular dependence of MAE
on crystal structure parameters $\alpha$ and $\tau$
for two concentrations, $p=0$ and $p=10^{21}$~cm$^{-3}$.
The impact of the lattice parameters is qualitatively
different in the insulating and conducting crystal.
The change in the internal displacement from $\tau=0.03$ to $\tau=0.02$
(continuous and dotted lines in Fig.~\ref{fig5}, respectively)
results in significant change of the amplitude of MAE in
the insulating case, while in the conducting case it is less
important.
On the other hand, changes in the rhombohedral angle $\alpha$, 
which are not
very important in the insulating crystal, produce much bigger 
effect in the conducting case. 
Similar differences are discussed in the next subsection, they
are related to the fact that
the spatial region around the Mn ion contributing to MAE is more
localized in the insulating case.

Let us also remind that \gmt\  containing more than
20~at.~\% of manganese is not rhombohedral but cubic,
and the easy axis of magnetization lies in plane of the layer
(i.e., it is perpendicular to the [111] direction).
Although our calculations were performed for much lower
manganese concentration and quantitative conclusions
cannot be drawn, the direction of changes in MAE values is
in accordance with experiment.

\subsection{Dependence of MAE on the geometry of the Mn surrounding}

\noindent
In the metallic cases, the calculated amplitude of MAE 
is at least an order of
magnitude larger than that observed in experiment.
However, one should keep in mind that the calculations were done
for supercells which represent a perfect translationally invariant
infinite lattice.  As it was mentioned in the Introduction,
in real mixed crystals there is always a microscopic disorder,
as a result of which each Mn ion is in a different surrounding.
Moreover, the experimental Mn concentrations are an order
of magnitude higher than those considered here, and thus
the Mn-Mn coupling must be taken into account.

\begin{figure}
\includegraphics*[width=\linewidth,height=0.66\linewidth]{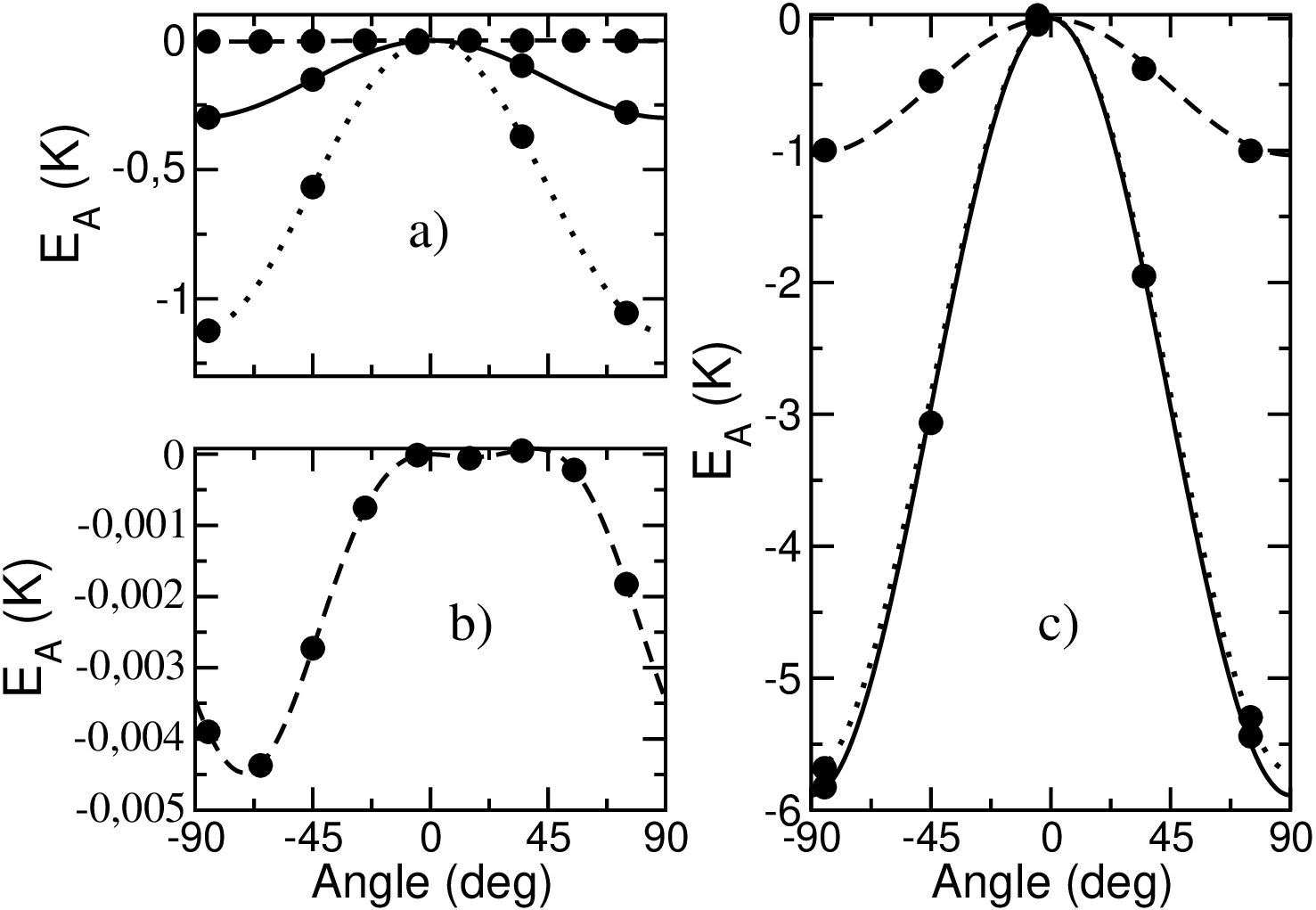}
\caption{%
Impact of the lattice parameters on the angular dependence of MAE for
the $2\times 2\times 2$ cell for (a), (b) $p=0$, NIP=512  and for (c) 
$p=5\times10^{21}$cm$^{-3}$, NIP=2744. The meaning of continuous, dotted and
broken lines is explained in the text.}
\label{fig6}
\end{figure}

Figure \ref{fig6} shows the angular dependence of MAE for 3 atomic
configurations of the Mn neighbours.
In all cases, the symmetry of the crystal is rhombohedral. The two left
panels, Figs.~\ref{fig6}a and \ref{fig6}b, show MAE for the insulating case, and
Fig.~\ref{fig6}c for
the hole concentration $p=5\times10^{21}$~cm$^{-3}$.

First, the continuous lines represent MAE
for the ideal lattices, characterized by $\alpha=88^{\circ}$
and $\tau$=0.03, and they already were presented in Fig. \ref{fig2}.
Due to the non-vanishing internal displacement $\tau$, 3 out of 6 Mn-Te
distances are equal 3.189 \AA, and the 3 remaining  are 2.819~\AA.

Second, the dotted lines show the results
for the configuration when the Mn ion is moved along the main
diagonal of the supercell by $a_0\sqrt{3}\tau$, and 
all distances to the six nearest neighbouring Te ions are the same.
In the metallic case, the displacement of the Mn ion has
a very small influence on MAE, while in the insulating
case the impact is substantial.
More specifically, the absolute values of changes of
$E_A(\theta=-90^{\circ})$  are 0.2~K and 0.8~K in the
metallic and insulating cases, respectively,
while the relative changes are 3\% and 277\%, respectively.

This difference is connected with the much larger
spatial range and magnitude of spin polarization of valence
electrons in the metallic than in the insulating case.
This issue is discussed in detail in the following.
Consequently, in the metallic sample the spin polarization extends
over the whole supercell, and the very localized perturbation of the
lattice caused by a displacement of a single ion has a small influence on MAE.
On the other hand, in the insulating case the range of
spin polarization is much smaller, and therefore the system is sensitive
to the perturbation in the nearest neighbourhood of Mn.
That is why the changes of MAE are also larger.

Finally, the broken lines display the results obtained for configurations with
partially optimized atomic positions. In the optimization procedure, the Mn
ion together with its nearest 12 Ge and 6 Te neighbours were allowed
to move, while the remaining atoms in the supercell were kept
at their initial ideal positions.
The displacements of 19 atoms lead to a much larger perturbation
of the lattice than the displacement of the Mn atom only,
and the changes of MAE are significant in both metallic and insulating cases.
For this partially relaxed \gmt\  lattice, the Mn-Te bond lengths
are 3.043~\AA\ and 2.796~\AA.
Thus, although the replacement of Ge by Mn atom
does not lead to such a large local lattice deformation 
as in other IV-VI semiconductors 
(for example, in Pb$_{1-x}$Mn$_{x}$Te the difference between ionic radii
of Pb and Mn is much larger than that between Ge and Mn), 
the impact of local distorsions on MAE is certainly not negligible.

The details of MAE for $p=0$ and the partially relaxed lattice are shown in the
lower left panel of Fig. \ref{fig6}.
The broken line is a fit to the calculated points according to formula
resulting from group theory considerations
\begin{equation}
 E_A=a_2\cos 2\theta + a_{4c}\cos^2\theta\left(cos^2\theta -
7/6\right)\nonumber +a_{4s} \sin^3\theta\cos\theta +c,
\end{equation}
where the second and the third term on the right hand side are proportional to
the
angular dependent parts of the Legendre polynomials $P_4^0$ and $P_4^3$,
respectively. Only such terms are allowed for rhombohedral symmetry if we
consider terms up to fourth order in spin components. One can observe that such
a fitting reproduces well the calculated values. \\

\subsection{MAE of the Mn-Mn pairs}

We also performed preliminary calculations for $2\times 2\times 2$
supercells containing two Mn atoms.
For such supercells there are 21 possible nonequivalent
configurations. For most of them the symmetry is not rhombohedral any more,
thus 
 Eq.~(\ref{eq7}) is no longer valid, and the term
$b_2\sin 2\theta$  must be added to the fitting formula for $E_A$.
The coefficients $a_2$ and
$b_2$ depend on both the configuration and the hole concentration.
Interestingly, the average of $b_2$ over configurations
almost disappears, indicating that although for most configurations 
the crystal symmetry is not rhombohedral,
after the averaging procedure the functional dependence of MAE
is in accordance with the macroscopic rhombohedral symmetry.

The situation is similar to that considered in Ref.
\cite{lusak_ssc}, where the single ion anisotropy of
Mn in Sn$_{1-x}$Mn$_{x}$Te crystal was analyzed.
In this work, the chemical microscopic disorder caused by the
random placement of Mn ions in the lattice was taken into account.
For each Mn ion, the amplitude of MAE was much larger than that
observed in experiment, and the angular dependence did not
correspond to the $O_h$ symmetry because the nearest neighbourhood
of Mn ions had not such symmetry.
However, after the averaging procedure the average MAE was
reasonably close to the experimental values, and the angular dependence
was consistent with the macroscopic cubic symmetry of the crystal.

\section{Mechanism of magnetic anisotropy}
\subsection{Spin polarization}
Atomic spin polarizations $\pi$, defined in Section II,
was analyzed for the $3\times 3\times 3$ supercells for two hole 
concentrations, $p=0$ and $p=5\times 10^{21}$~cm$^{-3}$.
The polarization depends on the atomic species,
the distance from Mn, and on the hole concentration.

For $p=0$, the Mn polarization $\pi^{Mn}=4.76$, which
is close to $\pi^{Mn}=5$ for the free atom.
The values of $\pi^{Ge}$ and $\pi^{Te}$ are much smaller,
of the order of 0.01, and practically vanish (i.e., they are less
than 0.001) beyond the first Te and Ge neighbours.

In the metallic case, the spin polarization of Mn depends on the hole
concentration, and changes from 4.76 to 4.68 for the considered
concentration range.
The main effect induced by the presence of free holes
is the pronounced increase of $\pi^{Te}$ of the nearest Mn neighbours from 0.01
to 0.07.
Also the spatial range of spin polarization of Te anions is larger,
since it vanishes only for distances greater than 10~\AA\ from Mn.
On the other hand, spin polarization of Ge cations is similar
to that in the insulating case. 
These results are explained by observing that
in the metallic phase, Mn polarizes free holes, i.e., the states
from the valence band top. These states are mainly built up from the
atomic states of Te, which are polarized by Mn.
The contribution of Ge orbitals to the hole states is much smaller,
and so is the spin polarization of Ge.
As it is shown in the next subsections, the strength 
of the spin polarization 
and its range are the decisive factors that determine MAE.

\subsection{Dependence on the spin-orbit strength}

\begin{figure}[t]
\includegraphics*[width=\linewidth,height=0.67\linewidth]{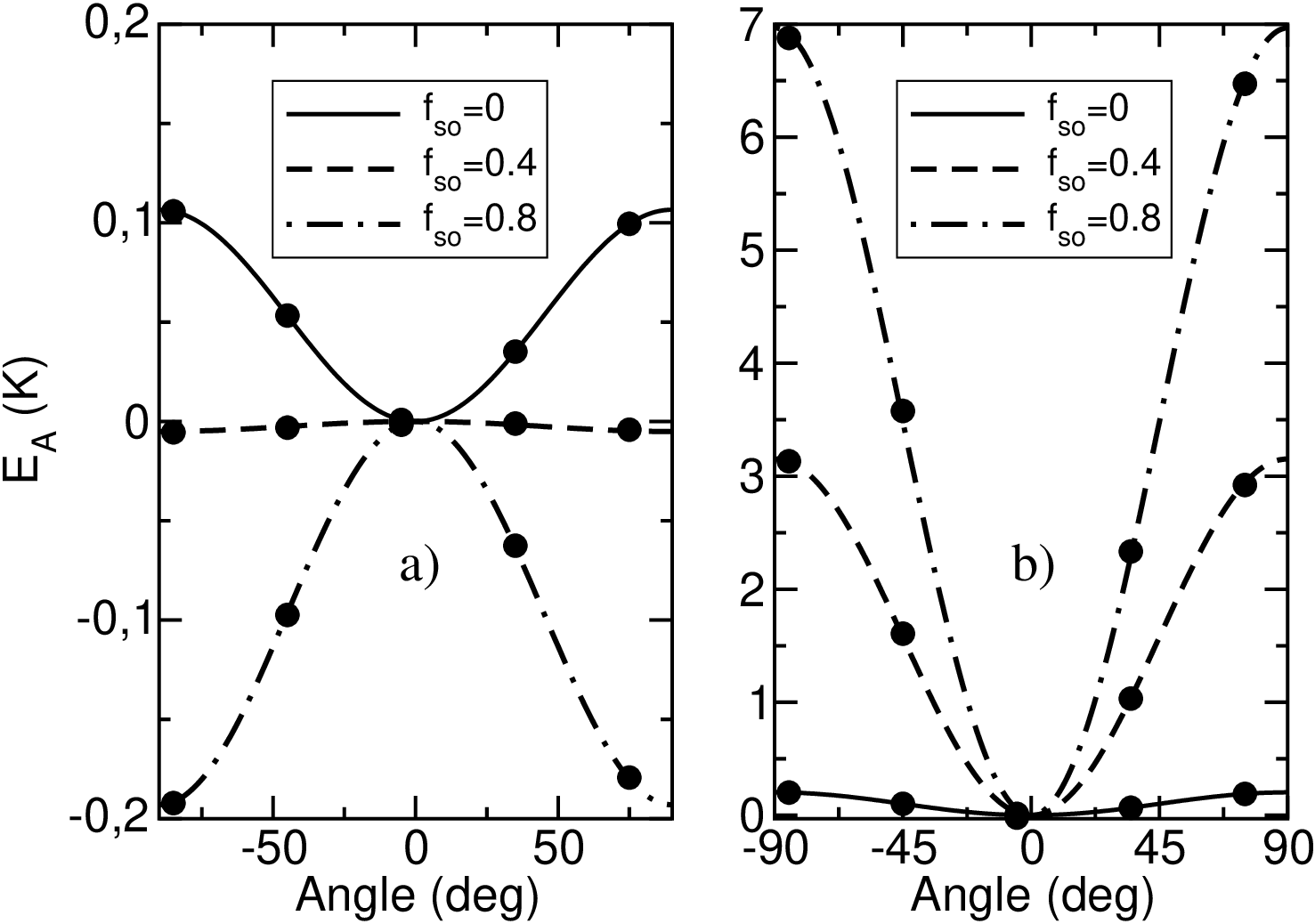}
\caption{%
Angular dependence of energy of magnetic anisotropy for a 
$2\times 2\times 2$ supercell with $\alpha=88^o$, $\tau=0.03$ containing 1 Mn
ion for hole
concentration (a) $p=0$  and (b) $p=0.5\times10^{21}$cm$^{-3}$.
The respective curves correspond to different S-O strength for Ge
and Te pseudopotentials.
}
\label{fig7}
\end{figure}

Figure ~\ref{fig7} displays the angular dependence of MAE for different
values of $f_{so}$, the factor that tunes the spin-orbit interaction strength
for Ge and Te.
The figure compares the results for the insulating $p=0$ case,
Fig.~\ref{fig7}a, with those for the metallic one with
$p=0.5\times10^{21}$cm$^{-3}$, Fig.~\ref{fig7}b.

When $f_{so}=0$, i. e., the S-O coupling is non-vanishing 
for the Mn ion only,
the angular dependencies of MAE are nearly the same
in the insulating and the metallic cases (note that the energy
scales for left and right panels of Fig.~\ref{fig7} are different),
and the anisotropy is very weak.
It follows from the non-vanishing S-O coupling of Mn, together with
the hybridization between its orbitals and the orbitals of neighbouring
atoms.

When the S-O interaction is turned on, the neighbouring atoms
contribute to MAE as well.
The reason is that for a given crystal structure, from the total
energy point of view, each atom has its preferred direction of the
polarization which, in general, is not the same as the actual direction
of the manganese spin. That is why a change of direction of Mn spin
changes total energy.

The impact of the increasing strength of the S-O interaction
is different in the insulating and the conducting systems.
In the former case, the anisotropy changes sign, but MAE remains
small.
In the metallic regime, the amplitude of MAE can increase by
an order of magnitude or more, and it can also change sign depending
on the actual hole concentration, see Fig. 3.
We also found that the polarizations of Ge and Te atoms and
their directions do not depend on $f_{so}$ significantly.
Actually, the differences in polarizations for Ge or Te
for different $f_{so}$ are less than 0.005.

These results clearly show that even in the insulating case
the magnetic
anisotropy is not related to Mn ion only but also to atoms in the Mn
neighbourhood, the
spatial range of which depends on the hole concentration.\\

\begin{figure}[h]
\includegraphics*[width=\linewidth,height=0.67\linewidth]{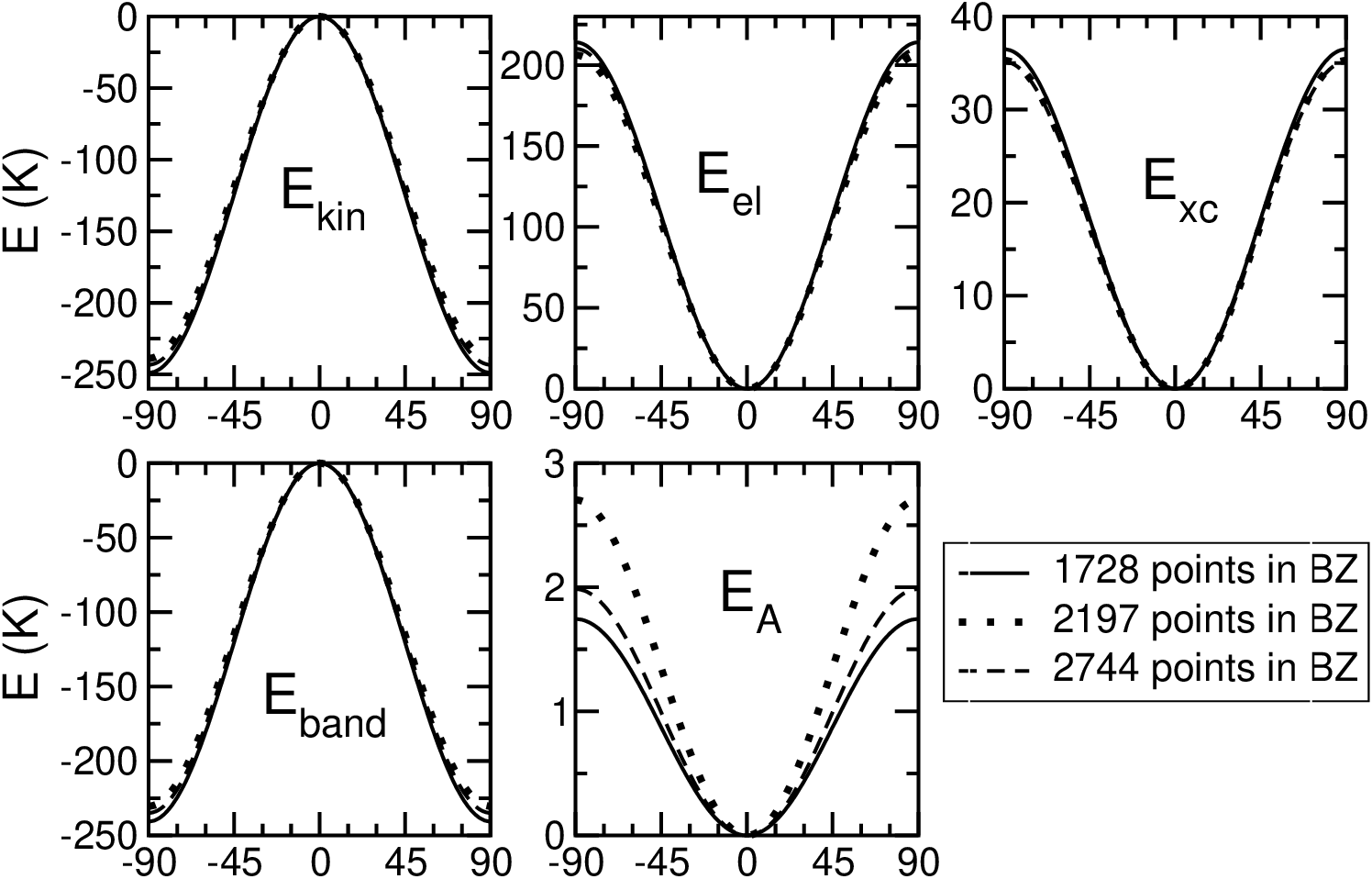}
\caption{%
Anglular dependence of energy of magnetic anisotropy components 
for a $2\times 2\times 2$ supercell, the hole concentration
$p=2.5\times10^{21}$cm$^{-3}$, $\alpha=88^o$, and $\tau=0.03$.
}
\label{fig8}
\end{figure}

\subsection{Components of the total energy}
Total crystal energy can be decomposed into
the contributions of kinetic energy,
electrostatic electron-electron and electron-core
interactions and exchange-correlation energy.
In Fig.~\ref{fig8} we show changes of the total energy, $E_A$, changes of
its components, $E_{kin}$, $E_{el}$, $E_{xc}$,  and also changes of the band
energy, $E_{band}$, i. e. the sum of Kohn -  Sham hamiltonian
eigenvalues over occupied states.
We note that the electrostatic term contains not only electron - electron and
electron - core interactions, but also the S-O interaction,
because the pseudopotentials are spin-dependent.
From Fig.~\ref{fig8} it follows that the changes of
components of the total energy are much
larger than the changes of $E_{A}$, which is typical for DFT
calculations.
The results of all the considered cases are qualitatively similar
to those shown in Fig.~\ref{fig8}. They clearly indicate that
the changes of spin direction of the Mn ion are accompanied by spatial
redistribution of electron density.

\subsection{Microscopic picture of magnetic anisotropy}

The above considerations allow us to propose the following picture
for the mechanism of magnetic anisotropy.
Assume that we know the total energies of the system,
$E_1$ and $E_2$, for two directions of the Mn spin,
${\bf \hat{n}_1}$ and ${\bf \hat{n}_2}$, respectively.
As it was discussed above, the direction of the Mn spin
determines the spin directions of remaining atoms in the system. \\
Let us assume now that we know the solutions to the Kohn-Sham
problem for the first direction of the Mn spin, ${\bf \hat{n}_1}$,
and let us denote this set of eigenfunctions, corresponding to
$E_1$, by $\{\psi_1\}$.
Let us now change, using a unitary transformation, the directions
of spins in $\{\psi_1\}$ in such a way that after the rotation
the polarization of atoms correspond to the second direction
of Mn spin, ${\bf \hat{n}_2}$.
The total energy $E_1'$, calculated with the new set of
wave functions $\{\psi_1'\}$, in general, is different from
$E_1$. 
This unitary transformation does not change electron densities
$n_{\uparrow}$ and $n_{\downarrow}$ 
(which are the eigenvalues of electron density
operator at a given point, Eq. (\ref{eq2})) thus
it does not change the kinetic, electron-electron, electrostatic 
part of the electron-core, and the exchange correlation energies. 
The only term in the Kohn-Sham hamiltonian 
which does not commute with the transformation is that 
describing the electron-core interaction, because pseudopotentials 
are spin
dependent. Thus, the difference between $E_1$ and $E_1'$ is solely due to
spin-orbit interaction. 
\\
If we now take this new set of the wave functions, $\{\psi_1'\}$,
as a starting point for the selfconsistent procedure, at
the end we obtain total energy $E_2$ and the set of wave
functions $\{\psi_2\}$.
The directions of polarization  of atoms calculated for
$\{\psi_2\}$ are the same as for $\{\psi_1'\}$, however the
resulting spatial electron density change.

Thus the change of spin directions
is accompanied by changes in spatial electron distribution.
These changes are, of course, very small and may be neglected
in phenomena in which we are interested in energies of the order of 1 eV,
however they are critical for the present problem of magnetic anisotropy.
One should also notice that the spatial redistribution of electrons
indicates that the so-called magnetic force  theorem \cite{liech},
sometimes used in calculations of magnetic anisotropy
\cite{jansen,kota} is not applicable
in the present case. Selfconsistent calculations are necessary for
each direction of the manganese spin.

On the basis of the presented results we can state that 
the anisotropy in \gmt, particularly
in the conducting case, is certainly not a single Mn ion anisotropy.
This is a collective effect in the sense that all the atoms
in the supercell give contribution to the phenomenon.
For higher manganese content, as is in the cases studied in
experiment, this is also the collective effect. 
This conclusion is in accordance with the Zener model \cite{dietl_ohno1,
dietl_ohno2} where the magnetic anisotropy is also due to free carriers.
However, {\em ab initio} calculations show that the differences in occupancies
of band levels are not the only cause, the redistribution of electron densities
should be also taken into account.

\section{Summary}

Magnetic anisotropy energy MAE in the diluted ferromagnetic 
semiconductor
\gmt\  was calculated within the Local Spin Density Approximation.
This approximation is shown to be accurate enough to describe MAE effects
of the order of 0.1 Kelvin. Low concentrations of Mn were considered, in which
magnetic ions
are interacting only weakly.
The analysis included the impact of factors that determine MAE,
namely (i) the concentration of free holes,
(ii) the microscopic configuration of atoms in the surrounding of Mn,
(iii) the macroscopic symmetry of the crystal (which in the case of \gmt\ 
can be  either cubic or rhombohedral), and (iv) the strength of the S-O
interaction in valence band.
The main conclusions are as follows.

First, magnetic anisotropy is largely determined by the presence of valence band
holes.
The increase of the hole concentration leads to both the variation
of the MAE strength and to the change of its sign,
i.e., to the reorientation of the magnetization
easy axis from parallel to perpendicular to the [111] direction.
In particular, for hole concentrations of the order $10^{21}$cm$^{-3}$,
typical of \gmt, the calculated direction of the easy axis of magnetization 
agrees with that observed in recent experiments on \gmt\  epitaxial layers
deposited on
 BaF$_{2}$ (111) substrates (perpendicular magnetic anisotropy)
\cite{przyby,knoff1,
knoff2, fukuma1}.

Second, MAE is surprizingly sensitive to both the macroscopic crystal structure 
and the atomic configuration of the Mn neighbourhood.
For example, a change of the crystal parameters from
$\alpha = 88^o$ and $\tau=0.03$ to $\alpha = 89^o$ and $\tau=0.02$
reduces MAE by as much as one order of magnitude.
A comparable sensitivity is found for the dependence of MAE
on the local atomic configuration.

Third, the calculated angular dependencies of MAE
fully agree with those predicted by group theory.
For \gmt\  in the rhombohedral phase, the terms of the second order in Mn spin
components satisfactorily describe the calculated results.
This dominance of the quadratic terms is consistent with
the experimental results \cite{przyby,knoff1a}.
On the other hand, higher order terms cannot be neglected for cubic \gmt.

Finally, the role of the spin-orbit coupling was elucidated by
calculating MAE as a function of the S-O strength of the band states of host
\gmt\  crystal.
The S-O of the host is particularly important in conducting samples.

The obtained results are related to the fact that the spin polarization of the
host
atoms have a decisive impact on MAE. Since the states from the top of the
valence
bands are mainly composed from the $p$ orbitals of Te, they determine
spin polarization of the valence states.
The insulating and the metallic case differ by the spatial extension of
the spin polarization: in the former case only the first two coordination shells
are spin polarized, while in the latter case spin polarization extends over
several lattice constants. This varying degree of localization explains
a number of the calculated trends in MAE.
Moreover, in the presence of S-O interaction the changes in the Mn spin
direction induce changes in the electron density,
the impact of which on the total energy was discussed in the last section.

Regarding the comparison with experiment one notices that a description of MAE
in real samples containing 10-20 per cent of Mn must include Mn-Mn magnetic
coupling,
much stronger in metallic than in insulating samples.
Moreover, incorporation of the microscopic chemical disorder,
which implies in particular local deformations of the lattice in the vicinity of
Mn,
is necessary.
On the other hand, the orientation of easy axis (perpendicular magnetic
anisotropy) observed in \gmt/BaF$_2$ (111) layer is already
explained by the present results.
Although the calculations were performed for \gmt, they directly apply to
closely related IV-VI 
ferromagnetic semiconductors, Sn$_{1-x}$Mn$_{x}$Te and
Pb$_{1-x-y}$Sn$_{x}$Mn$_{y}$Te. The proposed microscopic
mechanism of 
magnetic anisotropy should also be applicable to other groups of diluted
ferromagnetic semiconductors.

\ack
The authors acknowledge the support from NCN (Poland) research projects
no. UMO-2011/01/B/ST3/02486 and no.
UMO-2011/03/B/ST3/02664. This research was supported in part by PL-Grid
Infrastructure.

%
%

\end{document}